# Aristotle, projectiles and guns

Stephen M. Walley[*]
SMF Fracture and Shock Physics Group, The Cavendish Laboratory,
J.J. Thomson Avenue, Cambridge, CB3 0HE, UK

**Abstract**
When guns were developed in Europe in the 14th century, the theory of projectile motion was not the one we are familiar with today due to Galileo and Newton but the one taught by Aristotle approximately 1700 years earlier. In addition to Aristotle's wide-ranging philosophical concerns, his theory arose from the observation in everyday life that if an object is moving something must be moving it. This idea works very well for the horse and cart but is puzzling if you apply it to a thrown stone or spear. Problems with Aristotle's theory of projectile motion were identified by one or two people between his time and the 14th century, particularly John Philoponus (6th century AD) and John Buridan (14th century AD). An archer or a spearman does not need a theory of projectile motion, just a great deal of practice. But once the gun was invented it became important to know what angle a barrel should be oriented at and how much propellant to use, particularly as gunpowder was expensive. However, for many years afterwards the manufacturing techniques used meant that cannonballs were a loose fit to gun-barrels making cannons both inaccurate and of poor reproducibility shot-to-shot. Also air resistance makes the trajectory both impossible to calculate and qualitatively similar to theories based on Aristotle's writings. It was not until Galileo and Newton worked on the problem that a better theory of ideal projectile motion was arrived at.

## 1. Introduction

In 1849 Bonaparte asserted that cannons were first mentioned in the records of Italian and French towns early in the 14th century and that the English used them at the Battle of Crécy in 1346 [1]. Before the invention of the gun, it was not necessary for soldiers or their commanders to understand the theory of projectile motion. All that was necessary was many hundreds of hours of archery practice [2]. This all changed with the invention of gunpowder [3-5], which allowed the projection of large iron or granite spheres (cannonballs) (figure 1) or small metal shot (figure 2) at much greater speeds than previously possible [6-10].

---

[*] email: smw14@cam.ac.uk

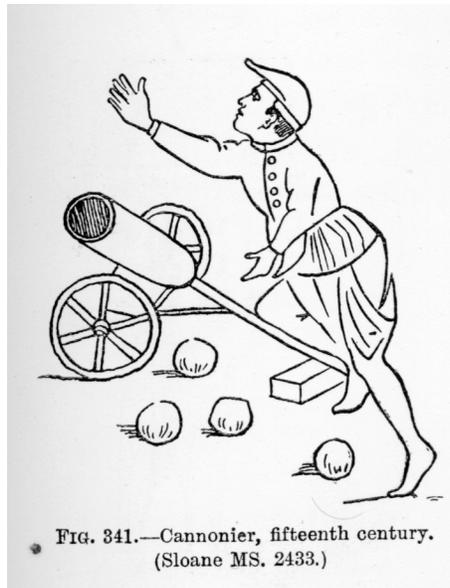

*Figure 1. Cannonier, fifteenth century. From [11] (page 263).*

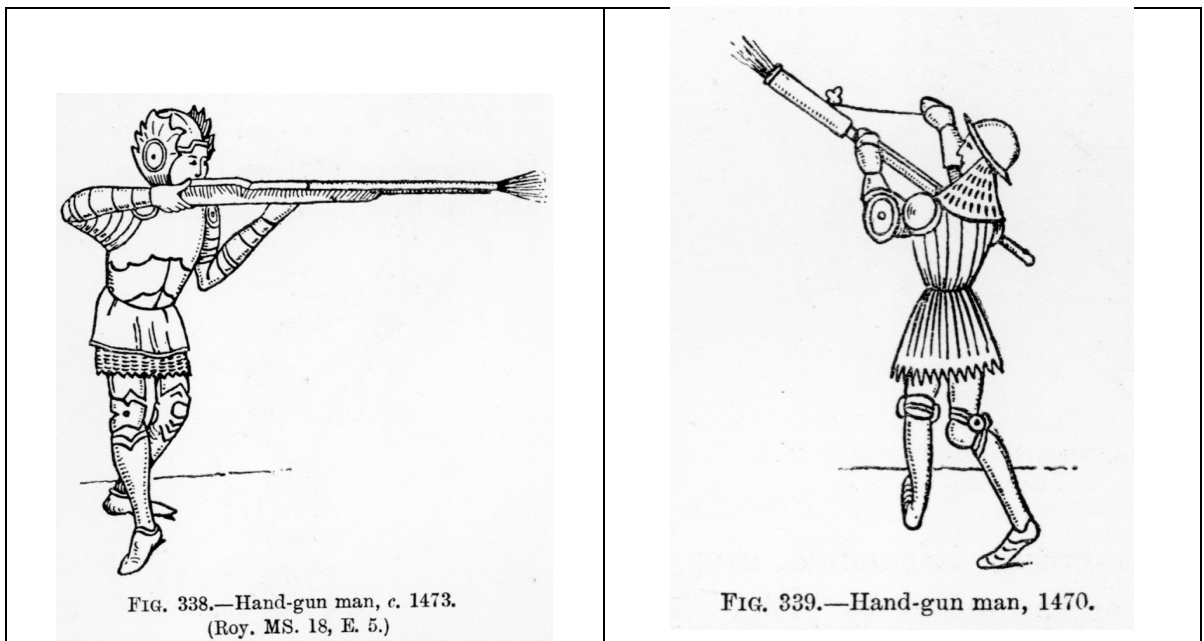

*Figure 2. Hand gun men 1470s. From [11] (page 262).*

J.D. Bernal pointed out in his book *Science and History* [12] the importance of cannon to the birth of modern physics. He wrote: "…the movement of the cannon-ball in the air (ballistics) was to be the inspiration for the new study of dynamics. … Impetus theory came long before the cannon, but the interest in the flight of the shot focused a new attention on it. The new mechanics differed from the classical in one vitally important respect: it depended on, and in turn generated, mathematics"

One unintended consequence of the invention of gunpowder was the breaking of the link between the desired effect and bodily effort and training which had been weakening since the invention of the counterpoise trebuchet in the 12th century (figure 3) [13-15]. So it became necessary to have some understanding of how to achieve the range required to, for example, fire a projectile over a city wall (figure 4).

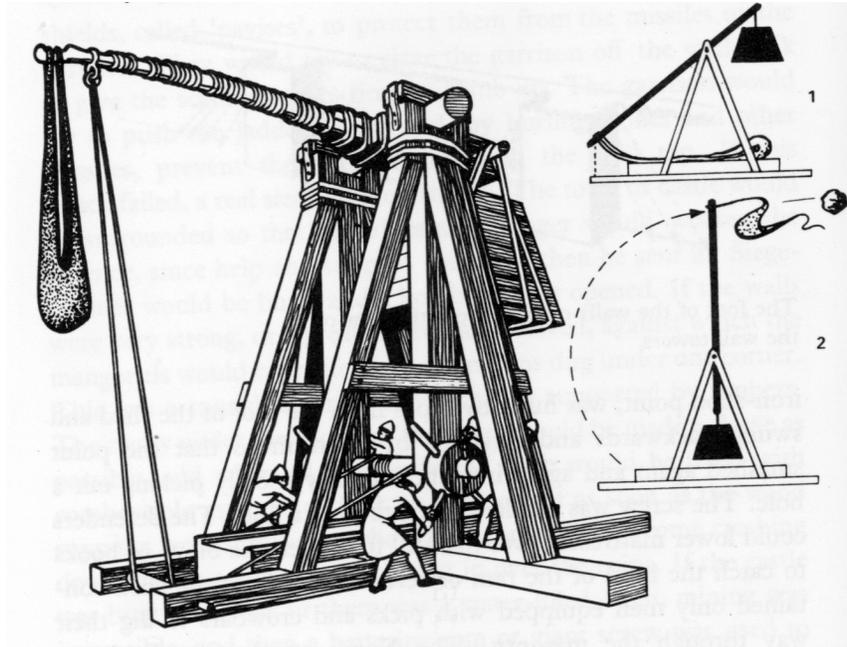

*Figure 3. Drawing of the counterpoise trebuchet and schematics of it in action. From [16] (page 56).*

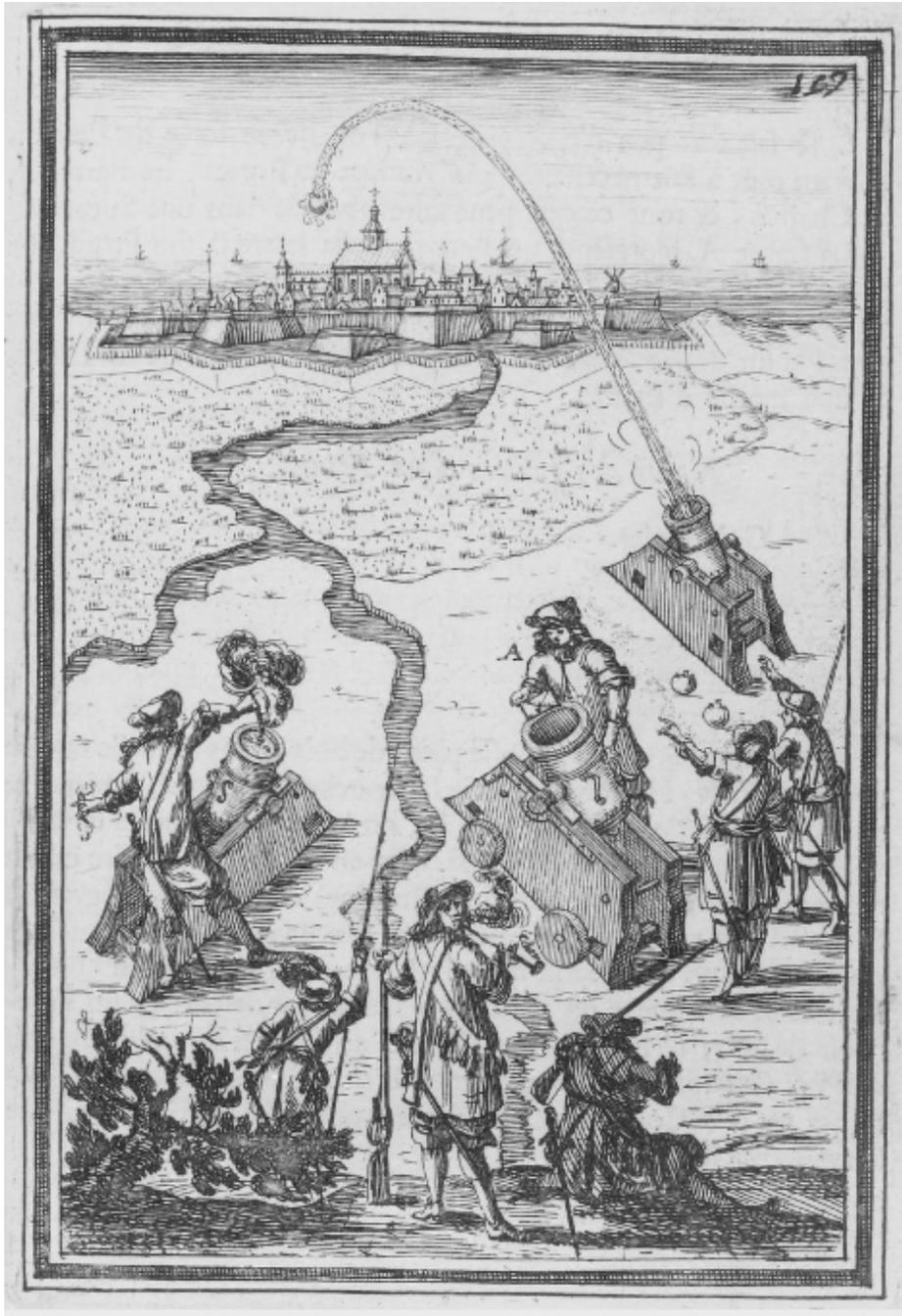

*Figure 4. Typical use of a mortar in the 17ᵗʰ century to fire over a city wall. From [17].*

## 2. Aristotle's theory of projectile motion

So what theory of projectile motion was available to the first gunners? One clue may be found in a diagram (figure 5) included in a set of handbooks (or magazines) published some centuries later (in 1669) to help English sea-captains do their job. At first sight, the trajectories appear qualitatively similar to what a modern calculation that takes air drag into account would predict (figures 6 and 7). However, closer inspection of the uppermost trajectory in figure 5 reveals that the motion is split into three distinct sections separated in the diagram by tick marks. From right to left, the initial, straight part is labelled 'The violent motion'. The second, curved part is labelled 'The mixt or crooked motion'. The third, vertically straight downwards part is labelled 'The naturall motion'. That this is not just an idiosyncratic English notion may be seen in the labels alongside the trajectories in a drawing published in 1592 in a Spanish artillery manual (figure 8).

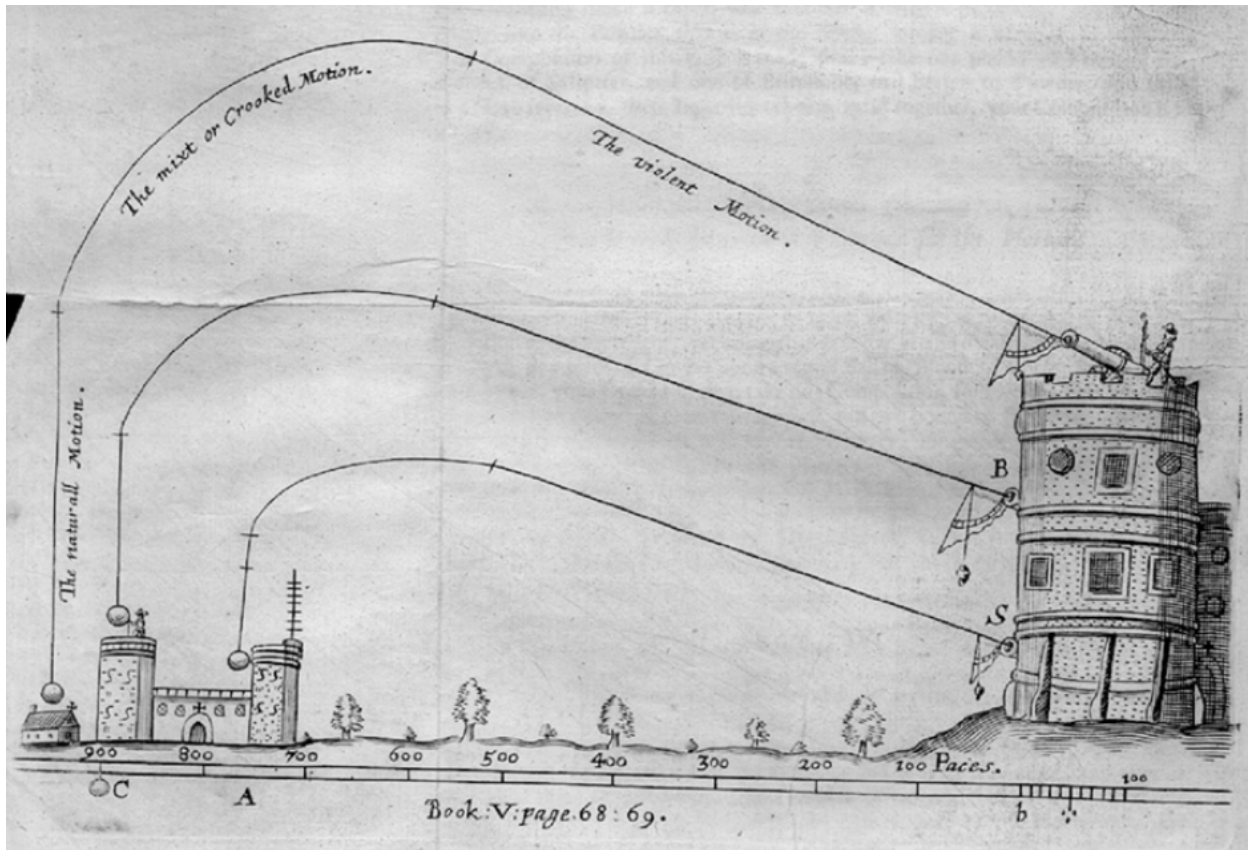

*Figure 5. Diagram of projectile motion in Samuel Sturmy's 'The Mariners Magazine. 5: Mathematical and Practical Arts' published in 1669. [18]. This image was recently discussed by Stewart [19].*

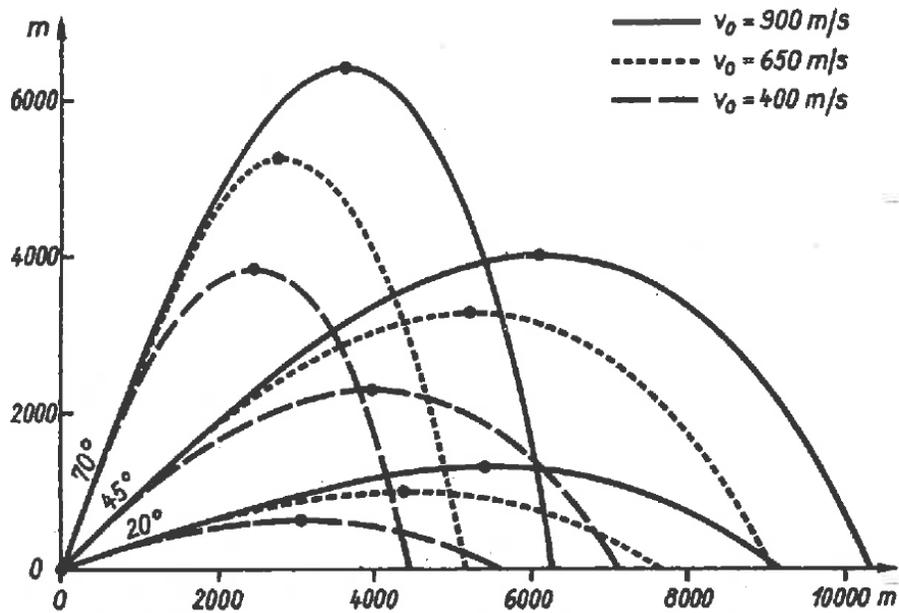

*Figure 6. Cranz's calculations of trajectories of projectiles fired at three different speeds and angles taking into account air resistance. Theory presented in [20]. Figure taken from [21].*

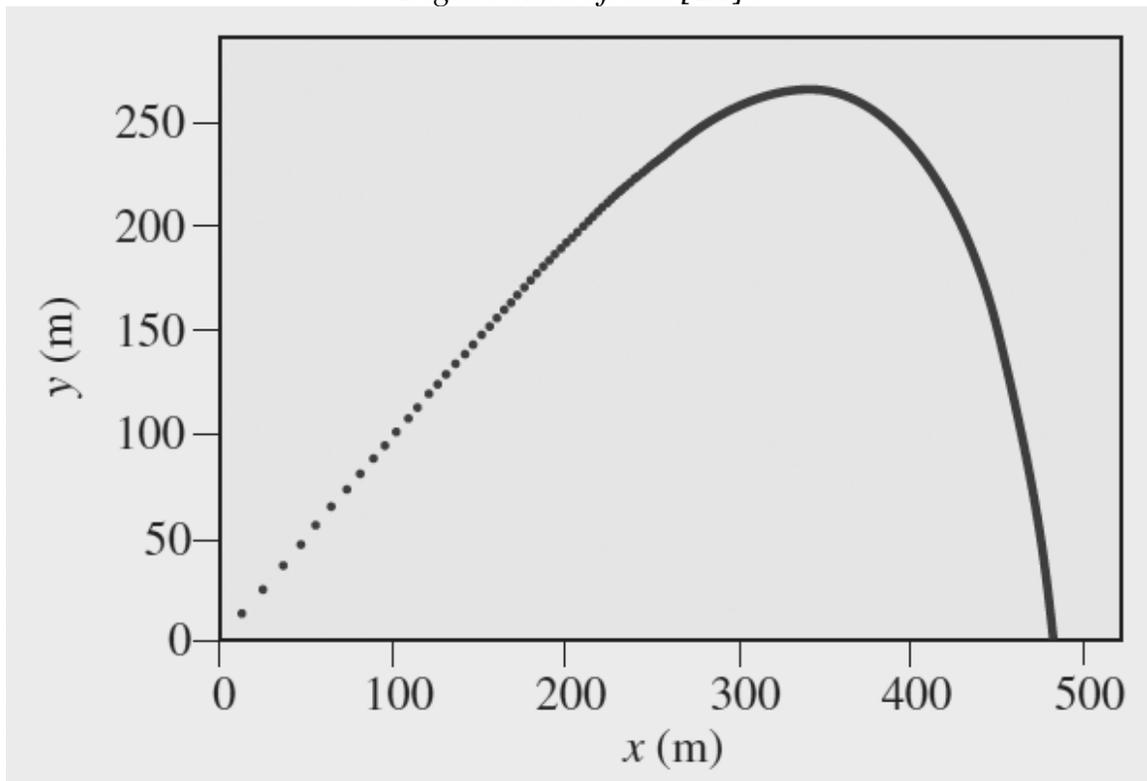

*Figure 7. Simulated trajectory of a cannonball assuming air drag force proportional to the square of the velocity. Initial speed 400 m/s, launch angle 45°. From [22].*

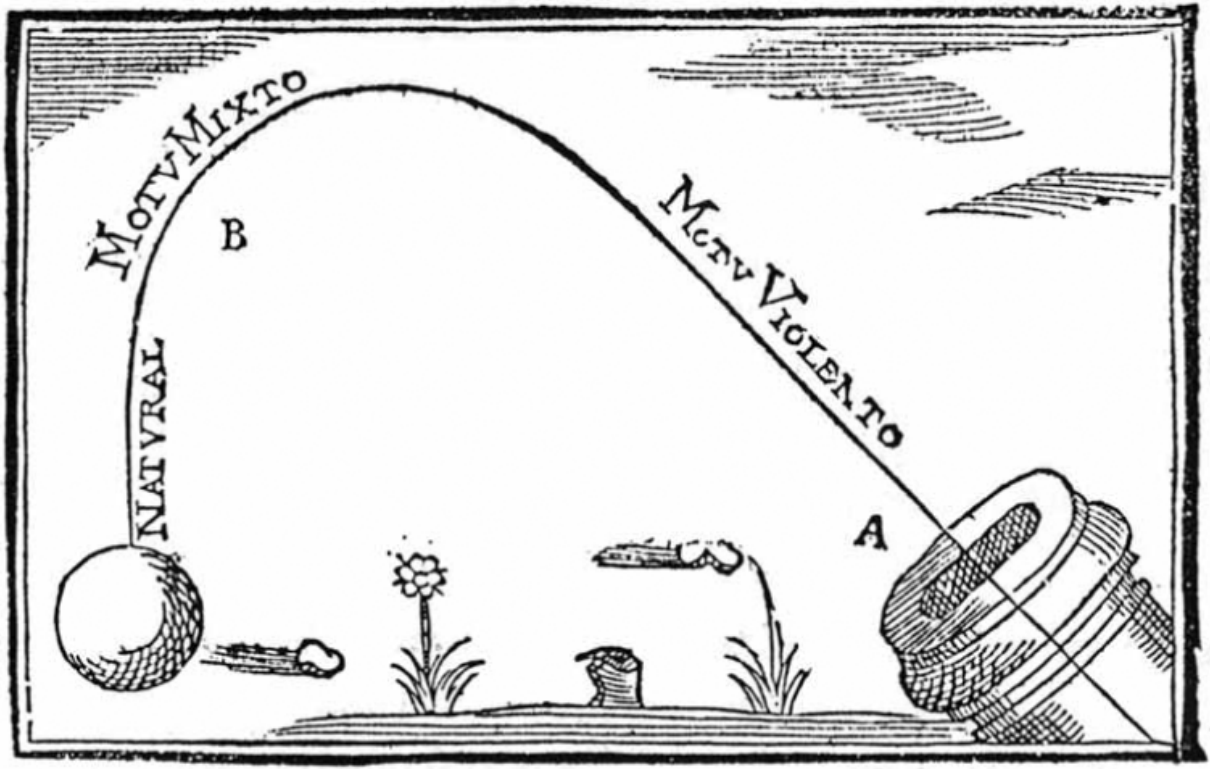

*Figure 8. Representation of violent, mixed and natural motion in a diagram taken from Luis Collado's 'Platica Manual de Artilleria' published in 1592 [23]. See also [24].*

Hannam commented in 2009 [15] that "Historians have long been puzzled how anyone could believe that a projectile could travel in a straight line and then drop out of the sky. After all, experience should have taught otherwise. But experience can be misleading. Bowmen were well aware that they could shoot straight at a target for maximum accuracy or fire into the air for maximum range. Those under a hail of arrows would have noted that they came from above and, under the circumstances, no one would have bothered to measure the exact angle of incidence. The trebuchet also propelled its rock into the air and, by the time this landed, it had lost a good deal of its forward momentum to air resistance. It would have appeared to those under attack that the projectiles were coming from above."

Note also that figures 4 and 5 show that a major interest of early gunners was the delivery of projectiles over a city wall rather than to batter the walls down (though they did this as well). As late as 1695, Edmund Halley, who was soon afterwards (1702) appointed Astronomer Royal, commented concerning this [25] that: "It was formerly the opinion of those concerned in artillery, that there was a certain requisite of powder for each gun, and that in mortars, where the distance was to be varied, it must be done by giving a greater or lesser elevation to the piece. But now our later experience has taught us that the same thing may be more certainly and readily performed by increasing and diminishing the quantity of powder, whether regard be had to the execution to be done, or to the charge of doing it. For when bombs are discharged with great elevations of the mortar, they fall too perpendicular, and bury themselves too deep in the ground, to do all that damage they might, if they came more oblique, and broke upon or near the surface of the earth; which is a thing acknowledged by the besieged in all towns, who unpave their streets, to let the bombs bury themselves, and thereby stifle the force of their splinters".

The first known representation of what a projectile does when it is fired out of a cannon (figure 9) goes back to a book entitled *Nova Scientia* by the Venetian Niccolo Fontana, usually known by his nickname Tartaglia (i.e. Stammerer), and first published in 1537. Note that Giovanni di Casali and Nicolo Oresme are believed to be the first men to plot graphs of one variable against another some time during the mid to late 1340s [26]. Tartaglia was credited by Hall as being the founder of the theory of gunnery [27].

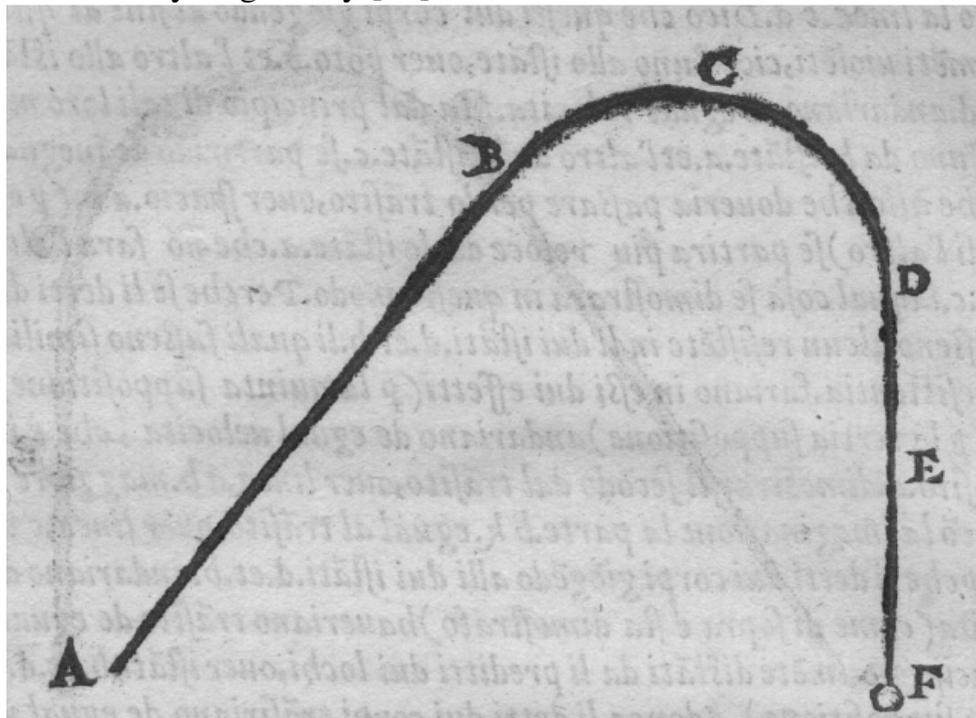

*Figure 9. Diagram of projectile motion as drawn by Tartaglia in 1537 (first published in English in 1588 [28]). AB is so-called 'violent motion', BCD is 'mixed motion', and DEF is 'natural motion'. Tartaglia's book 'Nova Scientia' has been translated twice into English during the 20th century [29, 30]. This graph was also discussed in a recent paper by Hackborn [31].*

It must be strongly emphasised that the physical theory behind Tartaglia's deceptively simple drawing is not that of Isaac Newton, who published his ideas some 150 years later in 1687 [32, 33]. Rather the physical theory available to Tartaglia had its origin in the writings of Aristotle (who lived 384-322 BC) [34], particularly in books 7 and 8 of his *Physics* and *On the Heavens* [35], albeit modified by people who grappled with his thought down the centuries. Again 'physics' had a different meaning in the writings of Aristotle to what it has today, namely the science of things as they are by nature [36].

The two commentators in in western Europe are usually held to be most relevant to the projectile problem (and whose writings would have been known to Tartaglia) are: John Philoponus (490-570 AD), who studied, taught and wrote in Alexandria [37], and John Buridan (1295-1363), who taught at the University of Paris [38, 39].

Before considering what Aristotle wrote about projectiles, we need to be aware that according to Barnes [40] and Grant [41], impressive and wide-ranging as Aristotle's surviving writings are, we have only about a quarter of what he is known to have written about [42], and what has come down to us is 'a compilation of his working drafts' rather than a body of finished texts. However, the thinkers who grappled with his writings down the centuries may not have known this, so that they would have approached his writings as a complete body of thought of which projectile motion was a small but very important part [43, 44]. Thus those whose interest was projectile motion often expressed perplexity at the apparent inconsistencies between what he wrote on this topic in his various books.

So what did Aristotle say about projectile motion? The issues that he raised and that his successors grappled with may be expressed briefly by three quotes: (i) "Everything that is in motion must be moved by something." (this first sentence of *Physics* Book 7 [45] may also be translated as "Everything that changes is changed by something"); (ii) "If everything that is in motion is being moved by something, how comes it that certain things, missiles for example, that are not self-moving nevertheless continue their motion without a break when no longer in contact with the agent that gave them motion?" (from *Physics* Book 8, chapter 10; [46]); (iii) "Nature is a cause of movement in the thing itself, force a cause in something else. All movement is either natural or enforced, and force accelerates natural motion (e.g. that of a stone downwards), and is the sole cause of unnatural motion. In either case the air is employed as a kind of instrument of the action … that is the reason why an object set in motion by compulsion continues in motion even though the mover does not follow it up." (From *On the Heavens*, Book 3, Chapter 2; [47]; for a more recent translation, see [48]).

In 1987 Wolff [49] summarised Aristotle's thoughts on the projectile problem in two separate statements: (i) "In Aristotle's thought, projectile motion requires a medium for two reasons: first the medium causes motion to continue, and second it terminates it. Water and air can do both because insofar as they are light they facilitate motion; insofar as they are heavy, they hinder it." (ii) "Aristotle intended to use his theory of projectile motion to maintain a principle which was, in fact, incompatible with the transmission of force, namely the principle that 'anything moved is moved by something'. With regard to projectile motion, he applies this principle by saying that the moved object ceases to be moved at the same time as the motion of the moving cause comes to an end."

Other 20th century commentators have expressed a range of opinions.

Lang in 1998 [50] pointed out that Aristotle discusses projectiles within the context of a much larger and more ambitious discussion of motion. In Aristotle's thought, air is able from its own inner nature both to move and to be moved (Lang comments that Aristotle does not explain why). So while the stone is undergoing so-called 'violent' or unnatural motion the air moves the stone upwards but when the stone stops a sudden transition to 'natural' (i.e. vertically downwards) motion takes place during which the air accelerates the stone downwards (for the application of these ideas to gunnery in 1561, see figure 10). Thus initially according to Lang's interpretation of Aristotle, when the stone leaves the thrower's hand, the stone is 'handed over' to the air which then becomes the mover, or more accurately a succession of movers. According to Lang, the important thing for Aristotle was to identify what keeps the stone moving. (The lines shown enclosing the trajectory up to the point *k* in figure 10 are probably an attempt to visually represent this action of the air).

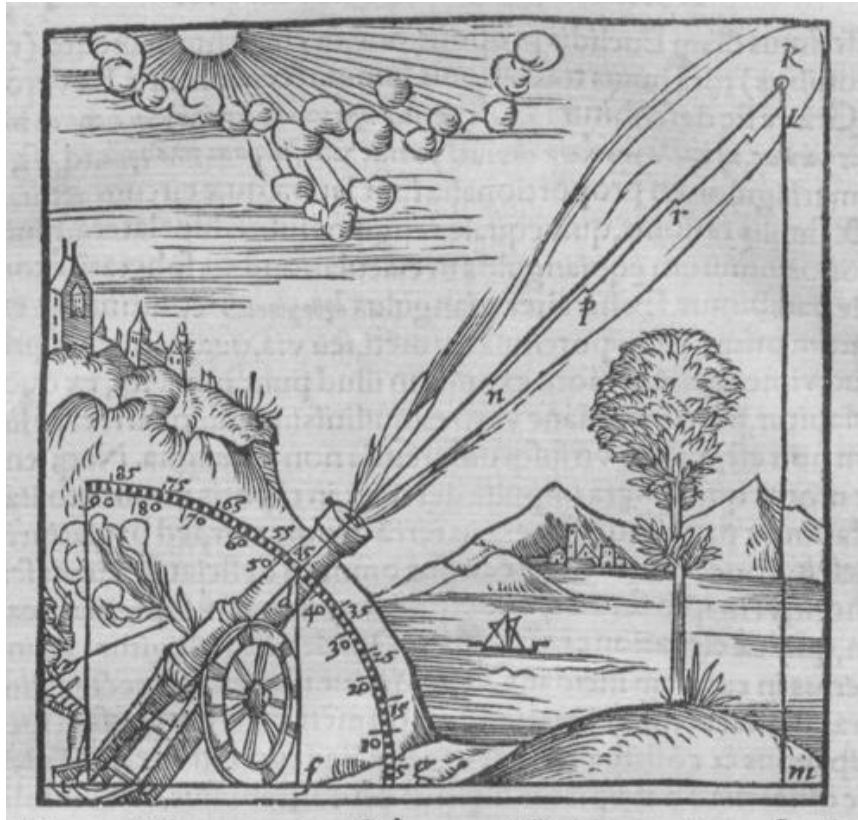

*Figure 10. Drawing by Daniel Santbech in 1561 of Aristotle's theory of projectile motion. From [51] (page 213). See also the recent discussion of these ideas by Stewart [19].*

Hussey commented in 1991 [52] (page 235) that Aristotle does not explain the mechanism of how the air acts to maintain the motion of the projectile. "The medium pushes in some way, that is pretty well all we are told."

Elders shed some light on this back in 1966 [53] by reminding his readers that Plato (*ca*. 425-348 BC) also discussed the projectile problem in his book *Timaeus*. Plato believed he had solved a number of interlinked problems by asserting that the air in front of the projectile is displaced and moves round to the rear where it pushes the projectile forward. Elders commented "Why does air have this astonishing faculty of moving while it is itself no longer moved, a faculty which the moving body is lacking?" Elders suggested that the answer may lie in connections Plato and Aristotle made between physics and biology (see also the more recent discussion by Franco [54]).

However, Ashley (writing in 1958) [55] said that Aristotle's overall aim was to try and obtain an understanding of nature as it is in itself: "If we are to understand a treatise such as *On the Heavens*, we must judge it in terms of what Aristotle was attempting to do. He hoped to construct a natural science which would rest on observation at every point". Ashley went on to point out that for Aristotle natural science was about the study of changes that occur in physical objects through natural processes, of which the most basic is change of relative place. He also asserted that for Aristotle "every body has a natural place, namely the one to which it is observed to move and where it achieves a stable condition".

This then is the origin of the concept of 'natural motion', by which is meant motion of a body towards its natural place. By contrast, 'violent motion' is motion of a body away from its natural place.

But the question may be asked: why does a thrown projectile transition from 'violent' to 'natural' motion i.e. stop moving upwards and start to fall? Hussey wrote [52] (page 235) that Aristotle identified two factors that slow the projectile. "First, some of the 'impulsion' has to be used up irreversibly in acting to divide the medium, so that the projectile can actually make progress. Second, in the reciprocal interaction [between the projectile and the medium] some power gets irreversibly dissipated."

John Buridan argued against the above and in favour of something called 'impetus' [56], a concept whose origins can be traced back to John Philoponus ($6^{th}$ century) [57]. According to Wolff, John Philoponus concluded that when somebody throws something they impart some incorporeal kinetic power to the object thrown rather than to the medium it is moving in. In other words, a hurled body acquires an 'inclination' or 'motive power' that secures its continued motion. This 'impressed virtue' is temporary and self-expending, so that all motion tends towards 'natural motion'. These ideas have also been discussed by Clagett [58], Sambursky [57], Franklin [59], Sorabji [60], Franco [54] and Graney [61]).

The following is based on the translation of some of John Buridan's writings by Marshall Clagett [62]. Buridan pointed out that there is no evidence 'in experience' for the air pushing projectiles along. Rather, air resists motion.

The three famous observations John Buridan made are: (i) millstones keep on rotating after they are disconnected from their driving mechanism even if shielded from air currents; (ii) javelins fly in the direction they are thrown even if they are equally sharp at both ends; and (iii) ships keep going up-river even after they have stopped being pulled. In the last case, he says "…a sailor on deck does not feel any air from behind pushing him. He feels only the air from the front resisting him."

According to John Buridan's understanding of impetus theory, a stone thrown upwards ceases to move and starts to fall down because the "impetus is continually decreased by the resisting air and by the gravity of the stone, which inclines it in a direction contrary to that in which the impetus was naturally predisposed to move it. Thus the movement of the stone continually becomes slower, and finally that impetus is so diminished … that the gravity of the stone wins out over it and moves the stone downwards to its natural place."

An amusing sketch illustrating these issues may be found on the first page of Pierre Varignon's *Nouvelles Conjectures sur la Pesanteur* (New Conjectures about Heaviness) published in 1690 (see figure 11). It shows two men (Mersenne and Petit) who have just fired a cannonball vertically into the air. The banner above the gun says in French "Will it fall back down again?". Note this drawing was intended to illustrate experiments performed by Marin Mersenne (the man on the left) in the 1630s [63], just before Galileo published his seminal work *Two New Sciences* in 1638 [64, 65]. Pierre Varignon began his book thus: "La premiére (*sic*) fois qu'on entend demander pourquoy un morceau de bois jeté en haut dans l'air, retombe toûjours; on pense avoir satisfait à la question, en disant: *C'est qu'il est pesant*. Et l'on ne crois pas qu'on puisse rien demander au delà." (The first time that you hear asked why a piece of wood thrown up into the air always falls back down again, you think you have satisfactorily answered the question by saying: *It's because it's weighty*. And you believe that's all there is to say on the matter).

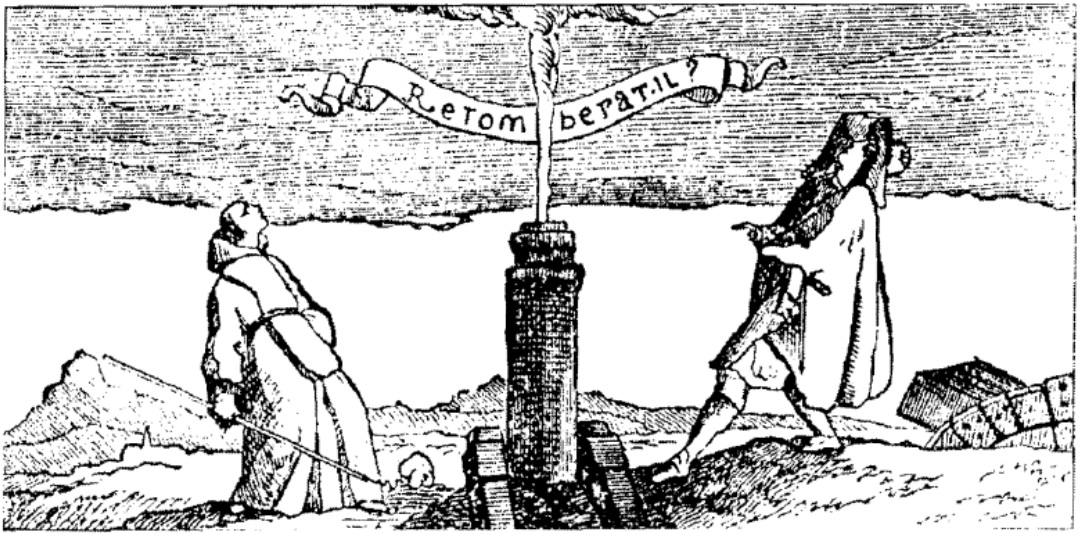

Figure 11. Will it fall back down again? From the opening page of "Nouvelles Conjectures sur la Pesanteur" by Pierre Varignon, 1690 [66] in which are reported experiments by Mersenne and Petit carried out in the 1630s. See also [63].

There are many problems associated with translating Aristotle's ideas on projectile motion into a modern language such as English, not least of which are that any words associated with motion, force, heaviness etc. will be interpreted by the reader through the prism of Newton's mechanics [59, 67]. It is also clear from the discussion above that Aristotle's ideas about projectile motion should be interpreted within the context of the whole body of his (and his predecessors') thought about nature [43, 68, 69]. But what is important to the matter at hand are the following questions: how did Aristotle's ideas on projectile motion result in the 16th century drawings of the trajectory of cannonballs fired from a gun (figures 5, 8-10, 12), and why does Tartaglia's sketch differ from those of Santbech and Collado?

## 3. The beginnings of the modern understanding of projectile motion

Why is a correct theory of projectile motion important? Figure 12 shows how Tartaglia interpreted the theory that had come down to him from Aristotle via Philoponus, Oresme and Buridan when thinking about how the angle of fire determines the range. The drawing shows that Tartaglia believed that the shot would fall straight down once violent motion had stopped (points E, D and K). Although he is often credited with introducing the non-Aristotelian idea of 'mixed motion', in fact he argued strongly against it as this quote shows: "I say that the mentioned body does not travel any part of its transit with a motion mixed of violent and natural motions, but travels only with a pure violent motion, or a part of it with a pure violent motion and another part with a pure natural motion. The instant at which the violent motion stops is the instant at which the natural motion starts. Assuming that the body could travel some part with violent and natural motions mixed together, which may be part CD, it follows that the mentioned body, while going from point C to point D, increases its velocity according to the ratio by means of which it shares a natural motion (because of the first proposition). Likewise, it decreases its velocity according to the ratio by means of which it shares a violent motion (because of the third proposition). It is absurd that the mentioned body increases and decreases its velocity at the same time." [30] (Tartaglia's first proposition was: "The farther each equally heavy body goes along its natural motion from its beginning, or the closer it comes to its end, the faster it travels" and his third proposition was: "The more an equally heavy body moves away from its beginning in a violent motion, the closer it gets to its end."). Reitan asserted that it was Albert the Great who came up with the non-Aristotelian idea of 'mixed motion' in the 13th century [70]. Smith has recently shown how the idea of 'mixed' motion was transferred across from optics [71].

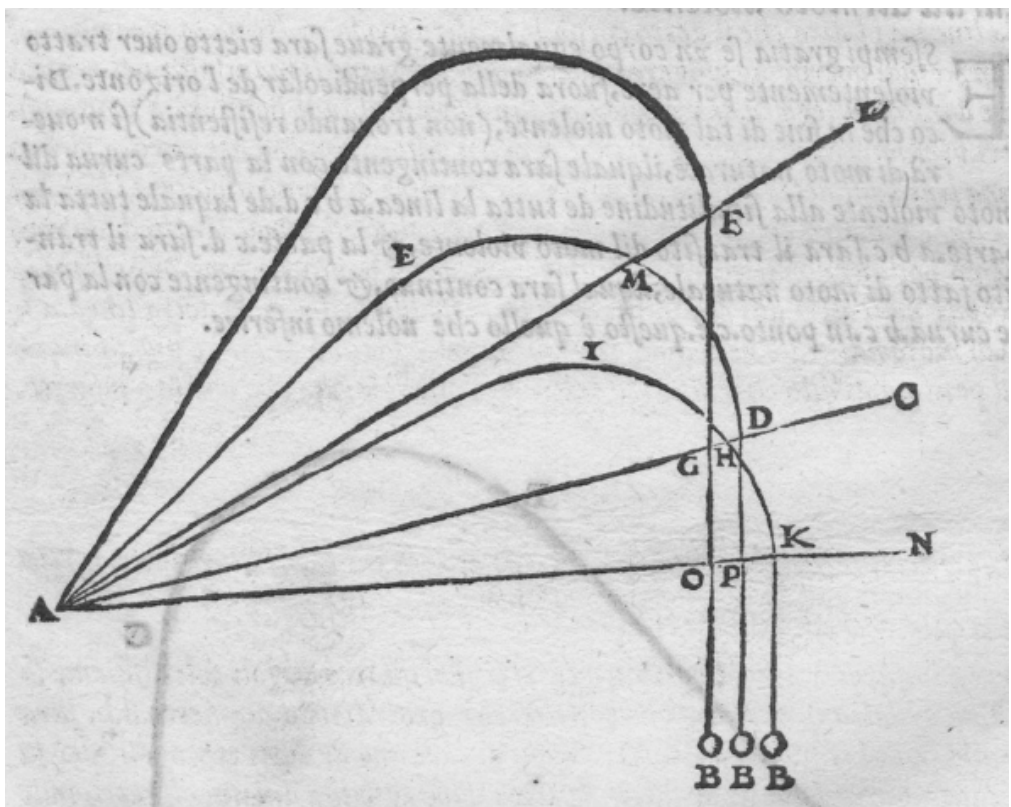

*Figure 12. Drawing showing how Tartaglia thought about the problem of working out what angle of fire would produce the greatest range. From [28].*

Tartaglia worked out theoretically that the maximum range of a gun would be obtained by angling the gun at 45° to the horizontal [72] [30] (page 69), although this is contrary to what the drawing reproduced here as figure 12 shows. Tartaglia went on to say that the friend who had asked him the question about the angle for maximum range thought that 45° was too large an elevation, but was convinced after some experiments were performed. However, Tartaglia's (unnamed) friend was correct: in a resistive medium (which, of course, air is), the angle for maximum range is less than 45° [73, 74].

It is also clear from the frontispiece to his book (figure 13) that Tartaglia was aware that the trajectory of projectiles fired upwards from a mortar was curved along its entire length. Indeed the curve shown looks qualitatively like a parabola. Note also in figure 13 a cannon to the right of the mortar is shown firing horizontally. It is clear that he knew that the shot does not hit the ground vertically. This observation is made even clearer in a book he published the following year (figure 14). These drawings make figures 9 and 12 even more puzzling. Tartaglia's mechanics was thoroughly studied and placed in its cultural context by Gerhard Arend in 1998 [75].

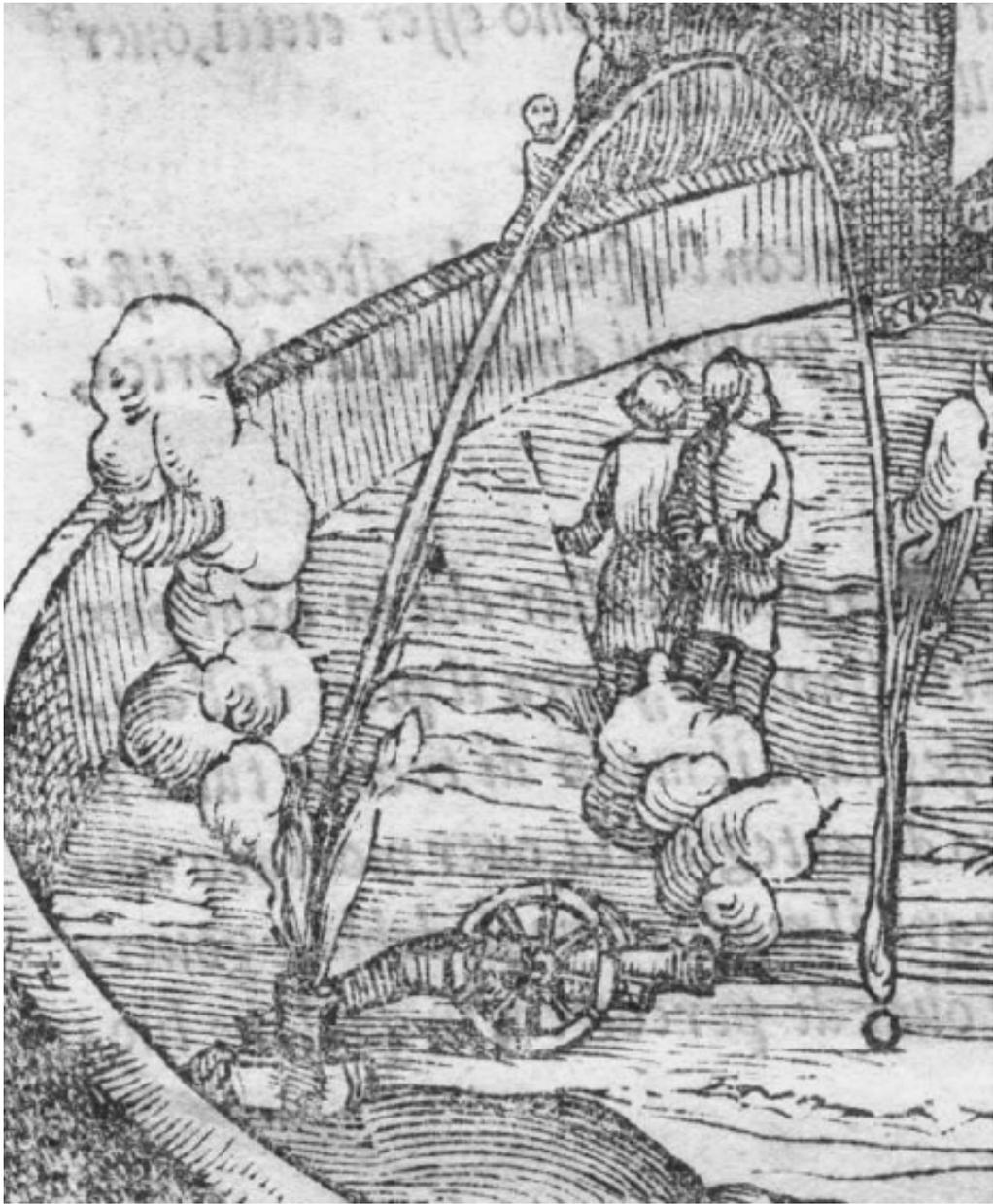

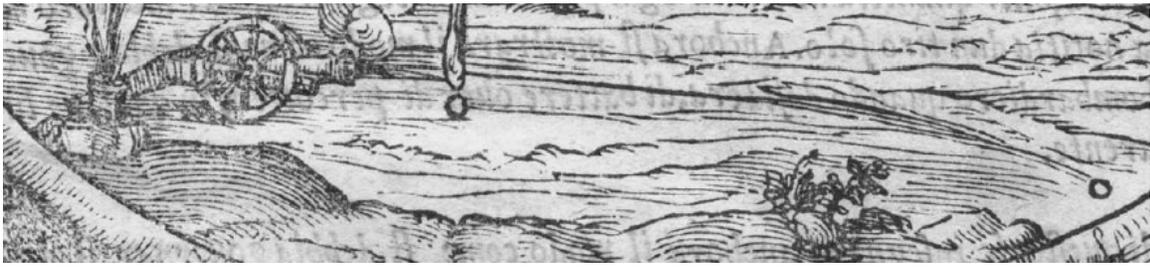

*Figure 13. Tartaglia's drawings from the frontispiece of his 'Nova Scientia' [28] of (a) the trajectory of a projectile fired upwards and (b) horizontally.*

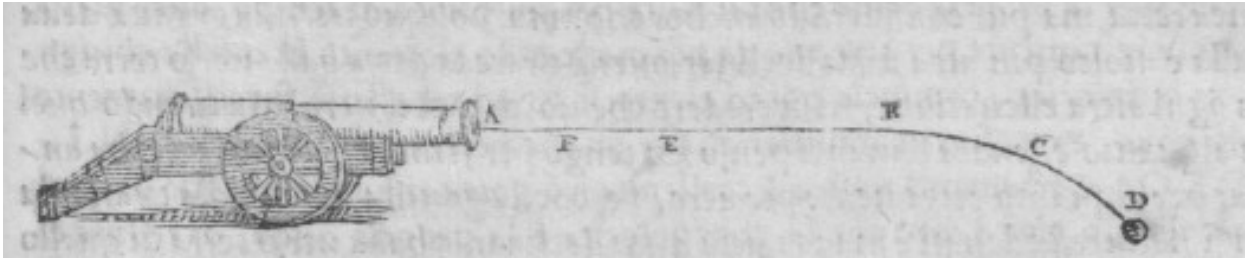

*Figure 14. Tartaglia's revised idea about projectile trajectories as set out in his 1538 publication 'Quesiti et Inventioni Diverse. 1' [76] (First English translation by Cyprian Lucar in 1588 [77]; for a more recent translation see [78]). See also the discussion by Walton in 1999 [79].*

Note that Galileo is usually credited [19, 80-82] with proving (in 1638) that the trajectory of a projectile is a parabola if it moves at a constant horizontal velocity and at the same time is uniformly accelerated in the vertical direction. Thus Galileo wrote in *Two New Sciences* (translation into English by Stillman Drake): "It has been observed that missiles or projectiles trace out a line somehow curved, but no one has brought out that it is a parabola. That it is … will be demonstrated by me…" [83] and "When a projectile is carried in motion compounded from equable horizontal and from naturally accelerated downward [motions], it describes a semiparabolic line in its movement" [84].

In 1979 Lohne found in the British Library some unpublished research by the Englishman Thomas Harriot (1560-1621) on ballistic trajectories (figure 15) [85]. Harriot took account of the gravity (or heaviness) of the shot, which causes it to deviate from the 'Aristotelian' straight line 'ad'. Like Tartaglia, the sketch shows that he erroneously believed that a projectile must hit the ground vertically. He also understood that air resistance results in a limit to the range, but did not explain how to determine the distance 'ac'. This is a difficult calculation even now [22, 86] and would have been beyond anyone's ability in the 17$^{th}$ century, although there were a few attempts in the 18$^{th}$ and 19$^{th}$ centuries [87-91]. Galileo pointed out the difficulties as long ago as 1638: "A more considerable disturbance arises from the impediment of the medium; by reason of its multiple varieties, this [disturbance] is incapable of being subjected to firm rules, understood, and made into science" [92]. However, Galileo went on to say: "In projectiles that we find practicable, which are those of heavy material and spherical shape, and even in [others] of less heavy material, and cylindrical shape, as are arrows, launched by slings or bows, the deviations from exact parabolic paths will be quite insensible" [93]. For more information on Thomas Harriot as a person and the studies he performed, see [24, 71, 94-98]. Late in the 19$^{th}$ century, Ingalls summarised the problem in his book entitled *Exterior Ballistics in the Plane of Fire*: "The molecular theory of gases is not yet sufficiently developed to be made the basis for calculating the resistance which a projectile experiences in passing through the air" [89].

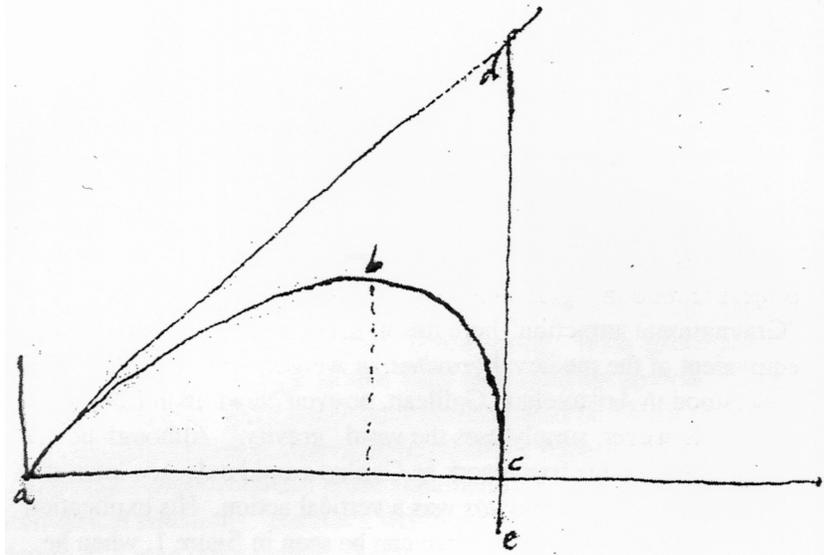

*Figure 15. Sketch by Thomas Harriot of the trajectory of the trajectory (abc) of a ballistic projectile under gravity. From [79]. See also the discussions by Lohne [85] and by Smith [71].*

**4. Projectile velocity measurements**

Knowledge of the initial projectile velocity is essential for ballistic trajectory calculations [22]. Once a description of gravity had been worked out by Newton and the acceleration due to gravity at the Earth's surface measured, a simple and straightforward way of determining projectile velocity would be to fire a gun up into the air and measure how far the shot ascends [99]. But air resistance means that this method gives a much lower value for the muzzle velocity than the true one. For example, Simmons writing in 1812 [100] (page 91) refers to a calculation by Daniel Bernoulli that air resistance is sufficient to reduce the height to which a cannonball might ascend from 58,750 feet (17,920m) to 7819 feet (2,380m).

One of the first articles reporting an investigation of the resistance of air to projectile motion was published by the Royal Society in 1687 [101]. But the first accurate measurements of air resistance to the motion of shot were performed by Benjamin Robins (1707-1751) and reported in his book *New Principles of Gunnery* [102]. He made the measurements by firing shot through a series of thin screens in order to determine the paths they followed. He found that the resistance of the air was not negligible, as was commonly believed at the time. For example, he calculated from the experimentally measured trajectories that the initial resistance of the air to a twenty-four pound cannonball, fired using sixteen pounds of powder, was about twenty-four times the ball's weight.

His experiments also showed that the paths that shot take through the air are neither parabolas (as Galileo believed) nor some other planar curve. In fact the balls usually followed trajectories with a double curvature, sometimes curving to the right and sometimes to the left of the direction of fire. He attributed this additional deviation to the spin imparted to balls as they rolled along the barrel of the gun due to their loose fit. He dramatically demonstrated this by bending the end of a musket to the left and firing it while it was firmly held in a vice. The expectation of the onlookers was that the shot would go to the left. In fact, the shot curved to the right. These experiments by Robins were also discussed in an article published in 1830 [103].

Another remarkable discovery that Robins made was the sudden large increase in air resistance when the shot is fired at around the speed of sound in air, known at that time to be just under 1100 feet per second (335 m s$^{-1}$).

In 1950, Corner credited Robins in the frontispiece of his book *Theory of the Interior Ballistics of Guns* [104] as being the founder of the study of interior ballistics of guns. Robins also studied the exterior ballistics of both rockets and guns. An assessment of Robins' investigations on ballistics was published by Johnson in a series of four papers in the *International Journal of Impact Engineering* and the *International Journal of Mechanical Sciences* [105-108]. Johnson's 1990 paper [106] set Robins' studies within the historical context of the development of the understanding of mechanics. One of the most remarkable illustrations that Johnson found dates from 1639 which shows a precursor of what is now called 'Newton's pendulum', the main differences being that cannonballs were used and the balls rested on a table rather than being suspended by strings (figure 16).

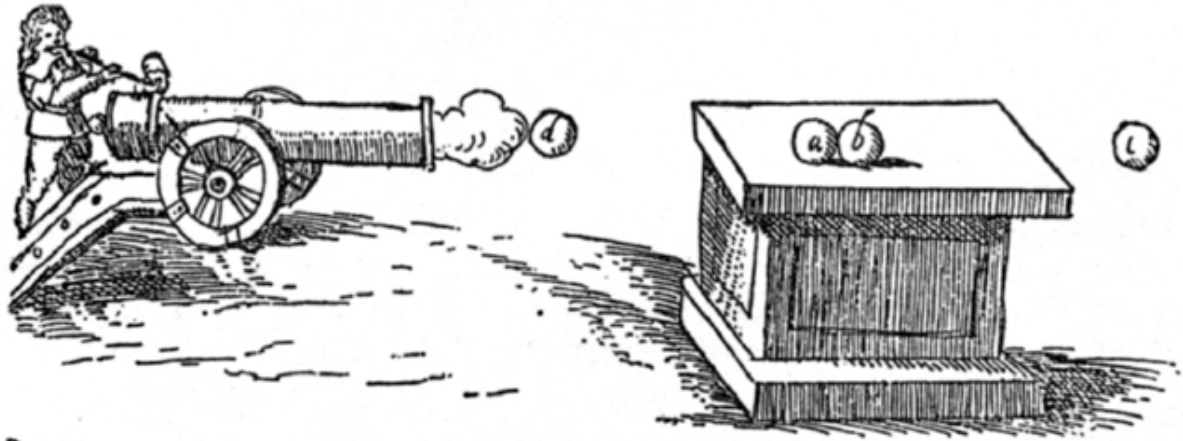

*Figure 16. Drawing showing the principle of transfer of momentum through a short chain of cannonballs. From [109].*

An assessment of the state of the art of gunnery before Robins began his investigations was published by Robertson in 1921 in his book entitled *The Evolution of Naval Armament* [110]. Robertson observed (page 114) that before Robins published his book *New Principles of Gunnery* in 1742 [102], the gunner was "primed with a false theory of the trajectory" (i.e. that due to Aristotle, discussed earlier in this article). In addition, balls in flight were believed to be affected by passing over water or over valleys. Furthermore, guns were imperfectly bored meaning that cannonballs were a loose fit so that they often "issued from the muzzle in a direction often wildly divergent from that in which the piece had been laid; on land it attained its effects by virtue of the size of the target attacked or by use of the *ricochet*".

Robertson continued (page 115) that: "The records of actual firing results were almost non-existent. Practitioners and mathematicians, searching for the law which would give the true trajectories of cannon balls, found that the results of their own experience would not square with any tried combination of mathematical curves"; (page 116): "For thoughtful men of all ages… the flight of bodies through the air had had an absorbing interest. The subject was one of perennial disputation. The vagaries of projectiles, the laws governing the discharge of balls from cannon, could not fail to arouse the curiosity of an enthusiast like Robins… Perusal of such books as had been written on the subject soon convinced him of the shallowness of existing theories. Of the English authors scarcely any two agreed with one another, and all of them carped at Tartaglia, the Italian scientist who in the classic book of the sixteenth century tried to uphold Galileo's theory of parabolic motion as applied to military projectiles. But what struck Robins most forcibly about all their writings was the almost entire absence of trial and experiment by which to confirm their dogmatical assertions. This absence of any appeal to experiment was certainly not confined to treatises on gunnery; it was a conspicuous feature of most of the classical attempts to advance the knowledge of physical science. Yet the flight of projectiles was a problem which lent itself with ease to that inductive method of discovering its laws through a careful accumulation of facts. This work had not been done."

According to Robertson, this state of affairs began to change in 1743 when Robins presented the findings of his book *New Principles of Gunnery* [102] to the Royal Society in London [111].

Thus on pages 117-118 of Robertson's book, we read: "In 1743 Robins' *New Principles of Gunnery* was read before the Royal Society. In a short but comprehensive paper which dealt with both internal and external ballistics, with the operation of the propellant in the gun and with the subsequent flight of the projectile, the author enunciated a series of propositions which, founded on known laws of physics and sustained by actual experiment, reduced to simple and calculable phenomena the mysteries and anomalies of the art of shooting with great guns. He showed the nature of the combustion of gunpowder, and how to measure the force of the elastic fluid derived from it. He showed, by a curve drawn with the gun axis as a base, the variation of pressure in the gun as the fluid expanded, and the work done on the ball thereby. Producing his ballistic pendulum [see figure 20] he showed how, by firing a bullet of known weight into a pendulum of known weight, the velocity of impact could be directly ascertained." Robins found that his theory (presented graphically in figure 17) agreed with his ballistic pendulum experiments (see figure 18) to within about 5%.

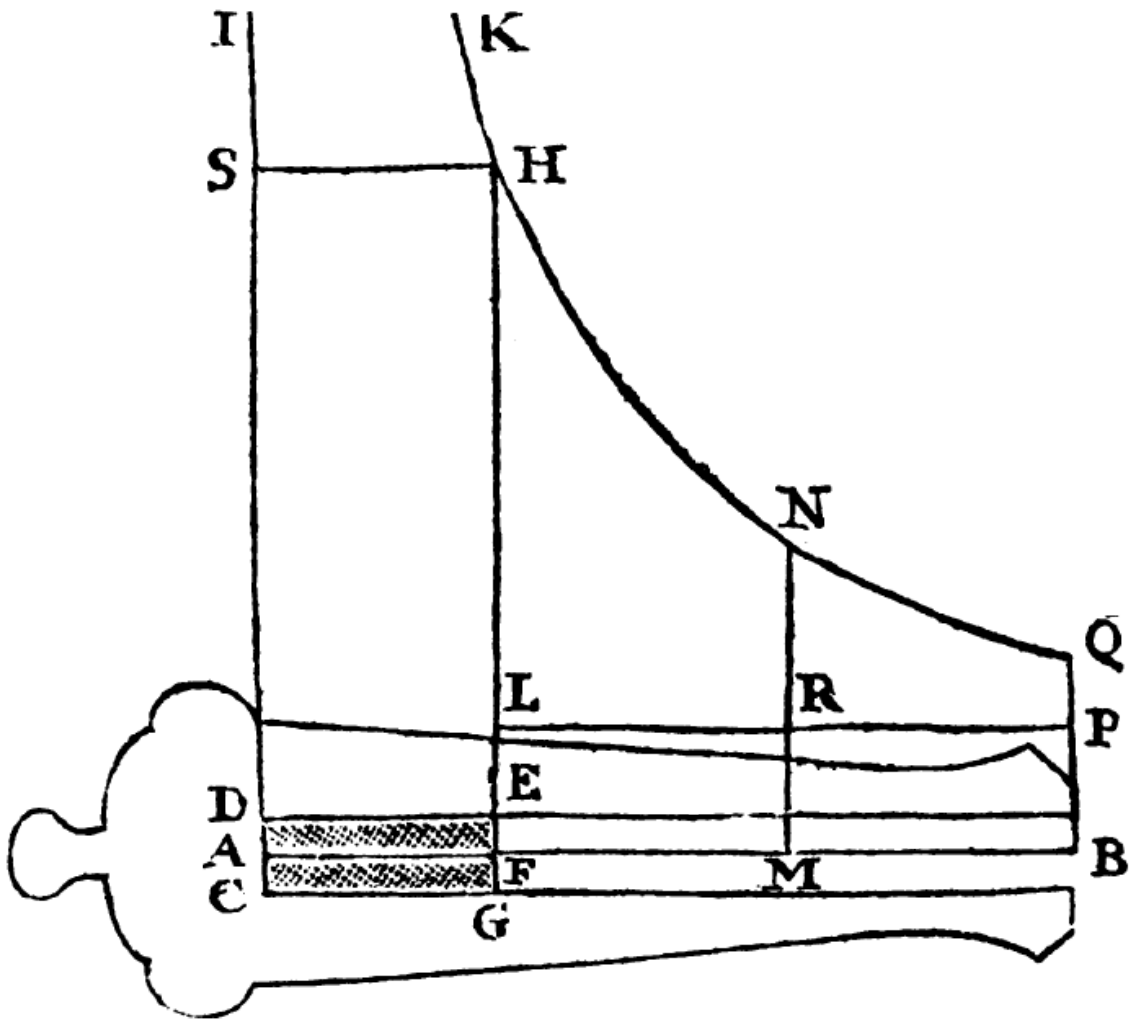

*Figure 17. Robins' 1742 graph of the variation of pressure in a cannon as the hot gases produced by the gunpowder push the shot along the barrel. From [102].*

According to Prony, writing in 1803 [112] (English translation published in 1805 [113]), the first measurements of projectile velocity were performed by Robins using a ballistic pendulum (figure 16) [102, 111]. Note that there was both a minimum and a maximum velocity that Robins' pendulum device could measure. The minimum, 400-500 ft s$^{-1}$ (120-150 m s$^{-1}$), was set by the requirement that the bullet be absorbed by the wood-facing on the pendulum bob (GKIH in figure 16). The maximum was set by requirement that the bullet does not penetrate right through the wood to the iron backing.

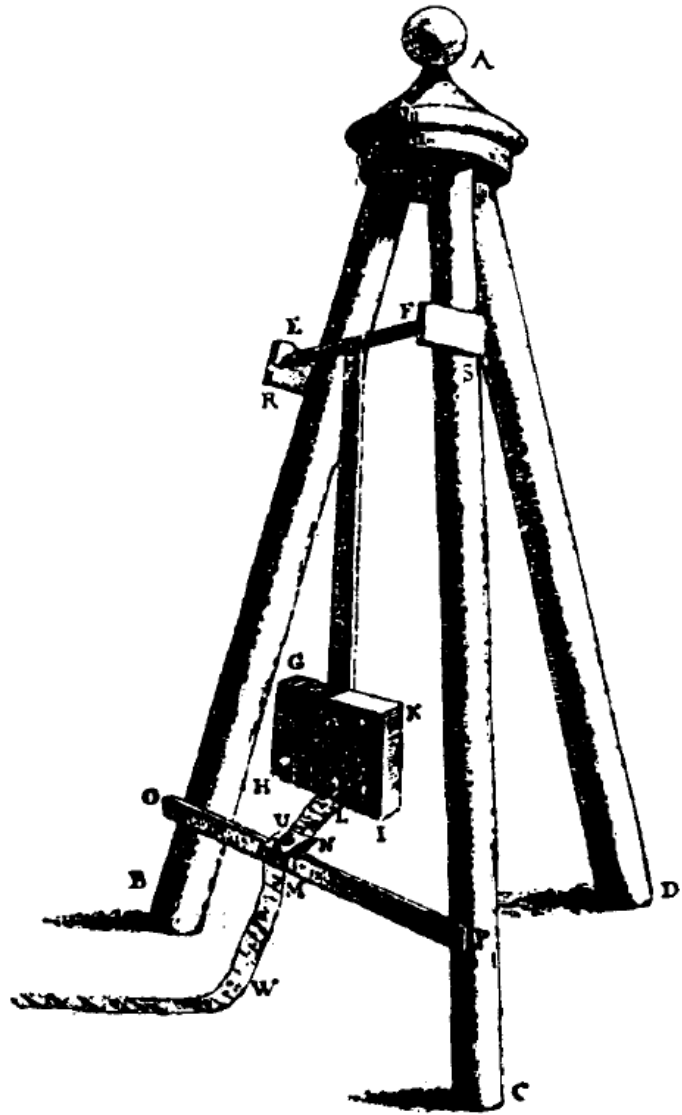

*Figure 18. Sketch of Robins' ballistic pendulum, used to make the first (indirect) measurements of projectile velocity. From [102].*

According to Prony, the most sophisticated measurements of projectile velocity at the time he wrote in 1803 had recently been made by Grobert. These experiments involved firing a shot through two spinning discs a known distance apart and which had also been engraved with lines at fixed angular spacings (figure 19). All that was required was to know how fast the discs were spinning.

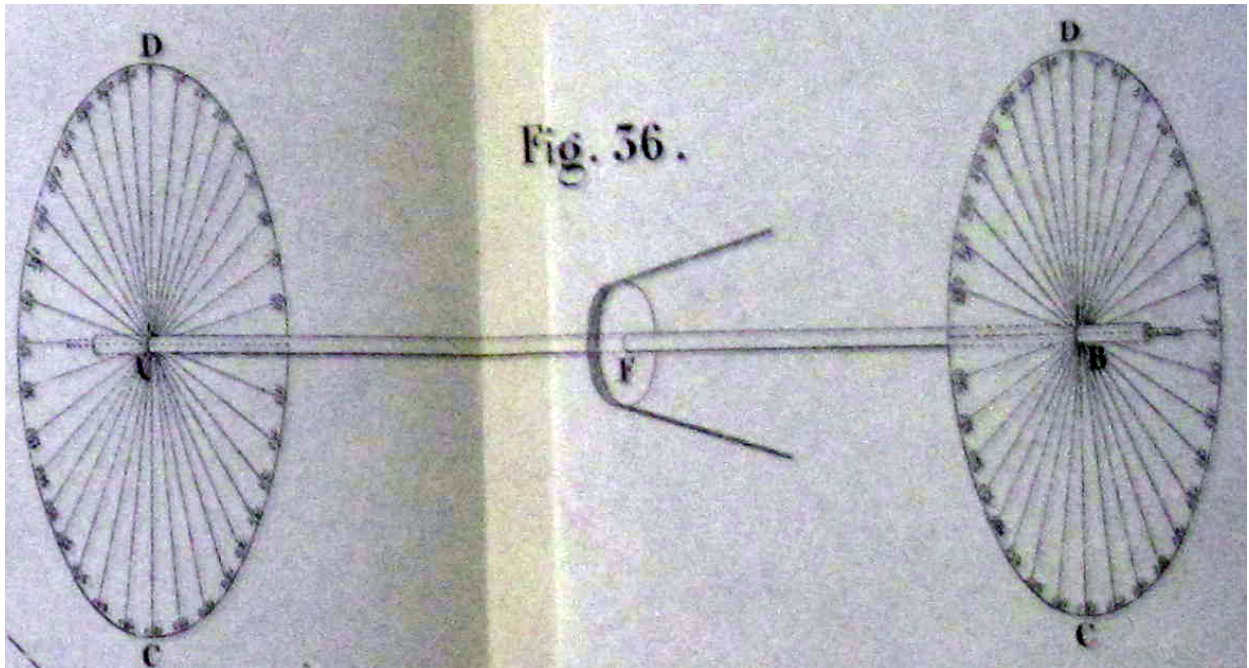

*Figure 19. Drawing of Grobert's spinning discs method of measuring projectile velocity directly. From [114].*

Prony described as follows how Grobert used the apparatus shown schematically in figure 19:

"A horizontal rotatory axis about 11 feet long carries at each extremity a pasteboard disk perpendicular to it, and fastened to it so that the whole may turn rapidly without deranging the respective positions of the parts.

A rotatory motion is given to the two disks be means of a weight suspended to the end of a cord, which, after having passed over a pulley ten or twelve yards from the ground, is rolled upon a wheel and axle level with the disks. An endless chain, passing round the wheel and the rotatory axis of the disks, communicates to this axis the motion which the weight in its descent imparts to the wheel.

The advantages this machine possesses over [its cylindrical predecessor] consist in the horizontal position of its axis, which admits the utmost degree of firmness and regularity in the position and motion of the disks: in the projectile not traversing a cylindrical surface, but two vertical planes, the extent and distance of which may be considerable, and this give very accurate measures: and its being capable, which no other apparatus is, of measuring the velocities of balls of different sizes projected at different elevations.

All that is necessary in using this apparatus is to give a uniform and known angular velocity to the disks; and to measure the arc comprised between two planes passing through the axis of the disk, and one of them through the hole in one disk, the other through the hole in its opposite.

In the trials made, the motion became sensibly uniform, when the weight arrived nearly in the middle of the vertical space it had to traverse, as was found by twice measuring the times of the third and fourth quarters of the descent, and afterwards comparing these times with the corresponding spaces passed through. An excellent stop-watch by Lewis Berthoud and another by Breguet, were used for this purpose. In most of the experiments the vertical space passed through by the weight was measured by the turns and parts of turns of the cord wound off in a given number of seconds, as in all respects most accurate and commodious.

To measure the arc a screen, or pasteboard, was fixed before each disk, a very little distance from it, and the hole in the first disk being brought opposite to the hole in its corresponding screen, a rod carried through the centre of these two holes and of the hole in the other screen which would be opposite them, must pierce the second disk in the plane of the hole in the first ; and the arc comprised between this point and the centre of the hole in the farther disk would measure the angle described by the apparatus while the ball was traversing the length of the axis.

It is obvious, that the fixed screens, which give the absolute direction of the path of the ball, afford the means of shewing the defect of parallelism, if there be any, between this path and the axis on which the disks revolve.

The gun-barrel was fixed horizontally, parallel to the axis of the disks, and at such a distance, that the concussion of the air by the explosion could not affect the motion of the nearest disk.

One thing may naturally suggest itself, that the time of the ball's passing from one disk to the other, through a space of three or four yards, must be less than $\frac{1}{100}$ of a second; and it is difficult to conceive, that in so short a space the disk could describe an arc capable of being measured.

But this difficulty is easily solved by the fact. When the motion became uniform, the wheel and axle commonly made 0.833 of a turn in a second; and every turn of the wheel produced 7.875 turns of the axis of the disks, which consequently made 6.56 turns in a second. Thus a point on the disk three feet from the axis would move about 41 yards in a second, and in $\frac{1}{100}$ of a second $\frac{41}{100}$ of a yard, or nearly 15 inches, a length more than sufficient for the most accurate measurement.

The experiments were made with a soldier's firelock and a horseman's carbine, the lengths of which in the bore were 3 ft. 8 in. and 2 ft. 5 in. The balls were accurately weighed, found to be on a medium 382 grains troy, and each was impelled by half its weight in powder.

The following formula was employed for calculating the velocity of the balls. Putting $\pi$ for the semiperiphery, when radius is unity = 3.141; $k$ for the ratio between the turns made by the wheel and axle and the arbor of the disks; $t$ the time employed by the wheel and axle to make a number of turns $n$; $r$ the distance of the hole in the second disk from the centre; *a* the arc described by this hole while the ball passes from one disk to the other; *b* the distance between the disks; and *V* the velocity of the ball; we shall have the equation $V = \frac{2\pi n}{kt} \cdot \frac{r}{a} b$.

The mean velocity deduced from ten experiments with the carbine was 1269 feet and a half in a second; that from the experiments with the musket, 1397 feet."

Prony goes on to describe other experiments performed by Grobert where the charge size was reduced, the effect of air resistance quantified and modifications made for performing experiments at various elevations up to 45°. Prony reckoned the apparatus could be enlarged to perform similar measurements for cannonballs, though how large the apparatus would have to be he said would have to be determined by trial. Grobert also apparently had automated the apparatus "to prevent any mistake from want of attention in the persons employed", but Prony drily remarks that "…complicated machinery is always liable to get out of order, and it may be dispensed with here, if the observers be ever so little expert and attentive."

## 5. Reflections

I was struck while researching this article how profound were the questions that Aristotle raised, but also how little of his thought is known by practising scientists nowadays, except for a few specialists and enthusiasts. This is largely due to the rubbishing of Aristotle by the English politician and philosopher Francis Bacon in his *Novum Organum* published in Latin in 1620 (for a modern English translation, see [115]). Examples of what Bacon wrote about Aristotle include: "Men become attached to particular sciences and contemplations either because they think themselves their authors and inventors, or because they have done much work on them and become most habituated to them. But men of this kind who apply themselves to philosophy and to contemplations of things in general, distort and corrupt them as a result of their preconceived fancies. The most striking example of this is seen in Aristotle, who utterly enslaved his natural philosophy to his logic, rendering it more or less useless and contentious" *Novum Organum* Book 1, Aphorism 55. "It was only in later times with the flooding of barbarians into the Roman Empire and the virtual shipwreck of human learning, that the philosophies of Aristotle and Plato, like light and insubstantial flotsam, survived the waves of time." *Novum Organum* Book 1, Aphorism 77.

## 6. Acknowledgement

The author expresses thanks to the late Professor W. Johnson for his interesting lectures on impact given at Cambridge in the late 1970s and early 1980s and for his set of publications on Benjamin Robins.